\documentclass[english,aps,prc,nofootinbib,superscriptaddress,twocolumn,letter]{revtex4}
\usepackage[latin9]{inputenc}
\usepackage{hyperref}
\hypersetup{
  colorlinks=true,        
  linkcolor=blue,         
  citecolor=cyan,         
}
\usepackage{breakurl}
\usepackage{graphicx}
\usepackage{epsf}
\usepackage{epsfig}
\usepackage{amssymb,amsmath}
\usepackage[usenames]{color}
\usepackage{amssymb}
\usepackage{times}
\usepackage{comment}

\newcommand\snn{\sqrt{s_\text{NN}}}



\begin{document}

\preprint{This line only printed with preprint option}

\title{External magnetic field induced paramagnetic squeezing effect in heavy-ion collisions at the LHC}

\author{Ze-Fang Jiang}
\email{jiangzf@mails.ccnu.edu.cn}
\affiliation{Department of Physics and Electronic-Information Engineering, Hubei Engineering University, Xiaogan, Hubei, 432000, China}
\affiliation{Institute of Particle Physics and Key Laboratory of Quark and Lepton Physics (MOE), Central China Normal University, Wuhan, Hubei, 430079, China}

\author{Zi-Han Zhang}
\affiliation{Department of Physics and Electronic-Information Engineering, Hubei Engineering University, Xiaogan, Hubei, 432000, China}

\author{Xue-Fei Yuan}
\affiliation{Department of Physics and Electronic-Information Engineering, Hubei Engineering University, Xiaogan, Hubei, 432000, China}

\author{Ben-Wei Zhang}
\email{bwzhang@mail.ccnu.edu.cn}
\affiliation{Institute of Particle Physics and Key Laboratory of Quark and Lepton Physics (MOE), Central China Normal University, Wuhan, Hubei, 430079, China}

\begin{abstract}
In non-central heavy-ion collisions, the quark-gluon plasma (QGP) encounters the most intense magnetic field ever produced in nature, with a strength of approximately 10$^{19\sim 20}$ Gauss.
Recent lattice-QCD calculations reveal that the QGP exhibits paramagnetic properties at high temperatures. When an external strong magnetic field is applied, it generates an anisotropic squeezing force density that competes with pressure gradients resulting from the purely QGP geometric expansion. In this study, we employ (3+1)-dimensional ideal hydrodynamics simulations to estimate the paramagnetic squeezing effect of this force density on the anisotropic expansion of QGP in non-central Pb+Pb collisions at the Large Hadron Collider (LHC). We consider both up-to-date magnetic susceptibility and various magnetic field profiles in this work. We find that the impact of rapidly decaying magnetic fields is insignificant, while enduring magnetic fields produce a strong force density that diminishes the momentum anisotropy of the QGP by up to 10\% at the intial stage, leaving a visible imprint on the elliptic flow $v_{2}$ of final charged particles. Our results provide insights into the interplay between magnetic fields and the dynamics of QGP expansion in non-central heavy-ion collisions.
\end{abstract}
\maketitle
\date{\today}

\section{Introduction}
\label{sec:1}

One of the most studied phenomena in heavy-ion physics at relativistic energies is the strong collective flow of the quark-gluon plasma (QGP), which originates from the relativistic hydrodynamic expansion driven by local pressure gradients.
For non-central collisions, the anisotropy of the QGP fireball enhances this collective flow, ultimately leading to a non-zero elliptic flow $v_2$ of charged hadrons observed at the Relativistic Heavy Ion Collider (RHIC) and the Large Hadron Collider (LHC). 
Additionally, a extremely strong magnetic field is generated as a result of the colliding charged beams traveling at velocities nearing the speed of light~\cite{Voronyuk:2011jd,Bzdak:2011yy,Deng:2012pc,Kharzeev:2012ph,Skokov:2009qp}. The effect of magnetic field in heavy-ion collisions was initially explored in relation to the chiral magnetic effect (CME)~\cite{Fukushima:2008xe,Kharzeev:2007jp}. Later studies also discussed the role of magnetic fields on chiral magnetic wave (CMW)~\cite{Kharzeev:2010gd,Wang:2021nvh,Wu:2020wem}, Magneto-Hydro-Dynamics(MHD)~\cite{Gursoy:2014aka,Inghirami:2019mkc,Nakamura:2022wqr,Peng:2023rjj,Mayer:2024kkv},  jet quenching~\cite{Tuchin:2010vs}, heavy flavor transport and the splitting of the directed flow~\cite{Das:2016cwd,Gursoy:2018yai,Sun:2020wkg,Jiang:2022uoe,Sun:2023adv}, the global polarization of $\Lambda/\bar{\Lambda}$ hyperons~\cite{Guo:2019joy,Peng:2022cya,Xu:2022hql} and the spin alignment of vector mesons~\cite{Li:2023tsf}.

\begin{figure}[htb!]
\includegraphics[width=7.5 cm]{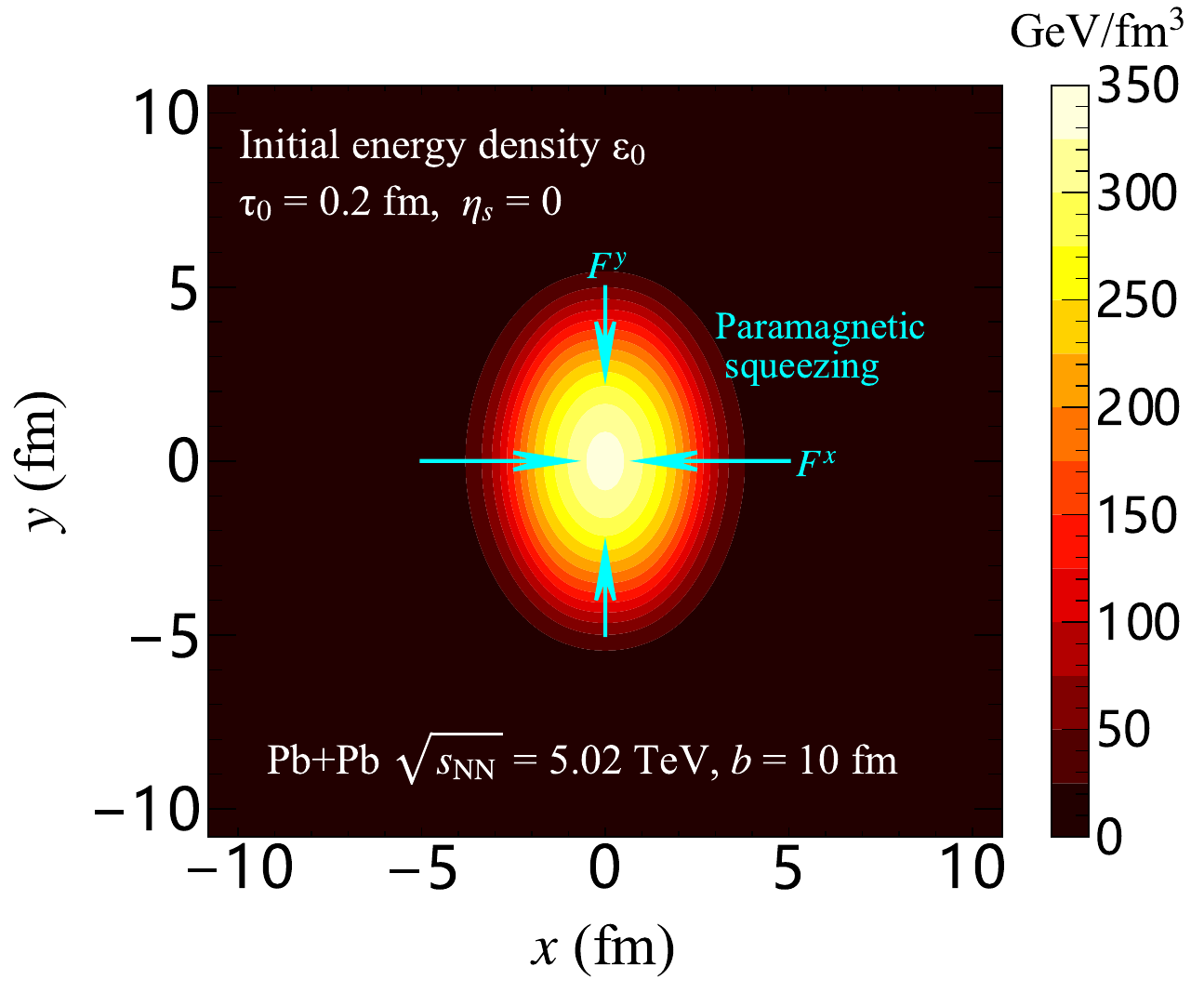}
\caption{(Color online) A sketch of paramagnetic compression in heavy ion collisions. The length of the arrow represents the magnitude of the force. The magnetic field induced squeezing effects cause QGP to be stretched in the $y$ direction.\label{Fig:energy}}
\end{figure}   

In this study, building upon previous studies by Bali et al.~\cite{Bali:2013owa} and Pang et al.~\cite{Pang:2016yuh}, we focused on exploring the response of the quark-gluon plasma (QGP) to external magnetic fields, also known as induced magnetization ($\mathbf{M}$), at the LHC energies. It is well known that when materials are exposed to an inhomogeneous magnetic field, paramagnetic materials align along the gradient of the magnetic field intensity $\lvert {\mathbf{B}} \rvert$ while diamagnetic materials align in the opposite direction~\cite{1984Electrodynamics}. The paramagnetic response in high-energy systems has been predicted through perturbation theory~\cite{Elmfors:1993wj} and the hadron resonance gas model in prior research~\cite{Endrodi:2013cs,Ding:2015ona}. Lattice Quantum Chromodynamics (QCD) calculations have revealed that at high temperatures, the QGP also exhibits paramagnetic properties~\cite{Bali:2013esa,Bali:2013txa,Bonati:2013lca,Bonati:2013vba,Levkova:2013qda,Bali:2014kia,Bali:2020bcn,Endrodi:2023wwf}.
In non-central nucleus-nucleus collisions, a strong inhomogeneous magnetic field with a strong spatial anisotropy is expected to generated~\cite{Deng:2012pc,Kharzeev:2012ph,ALICE:2019sgg,STAR:2021mii}. 
If the QGP encounters a non-uniform magnetic field and is driven towards regions of maximum $\mathbf{B}$ due to the positive $\mathbf{M}$, it will attempt to minimize its free energy to doing so, leading to a net force~\cite{Bali:2013owa,Bali:2020bcn}. This force density, referred to as ``paramagnetic~squeezing'' arises as the system strives to minimizes its free energy. The net squeezing force density can be expressed as  
\begin{equation}
\begin{aligned}
\mathbf{F} \equiv - \nabla{f} = \mathbf{(\nabla{B})\cdot{M}}.
\label{eq:F}
\end{aligned}   
\end{equation}
It was recognized that this force extends the distribution of QGP along the direction of the beam induced by the magnetic field, see Figure~\ref{Fig:energy}. In addition to the pressure gradient caused by the initial geometry, this squeezing force density elongation also affects the azimuthal structure of the system's expansion. Recently, by comparing the magnitudes of the squeezing force density with those of the pressure gradients at initial time for RHIC and LHC energies, several estimations of this effect have been made~\cite{Bali:2013owa,Pang:2016yuh}. The results suggest that this effect may be relatively small for RHIC collisions, but could be quite significant for LHC collisions. 

In this study, we detailed the estimate of the paramagnetic squeezing effect in various aspects at the LHC and High Luminosity (HL) LHC energy regions~\cite{Citron:2018lsq}. 
We employ the (3+1)-dimensional CCNU-LBNL-Viscous hydrodynamic model (CLVisc)~\cite{Pang:2018zzo,Wu:2021fjf} to simulate the time evolution of the energy density and the fluid velocity in the presence of a strong external magnetic field. The magnetic field is assumed to be given in terms of a few parameters that control the magnitude, the spatial distribution, the time evolution and the center-of-mass energy ($\snn$).  
By adjusting these parameters, we explore the impact of paramagnetic squeezing across diverse energy ranges at the LHC. The squeezing force density's impact is considered throughout the hydrodynamic expansion to investigate its influence on momentum anisotropy and to estimate its time-integrated effect on the $v_2$ of final charged hadrons. 

We also note that in this study, we follow Refs.~\cite{Bali:2013owa,Pang:2016yuh} and treat the magnetic field as an external degree of freedom, neglecting the QGP fluid's back reaction on the magnetic field.
A consistent description of the coupled evolution of hydrodynamic and electromagnetic field would require (3+1)-dimensional relativistic magnetohydrodynamics, which is still under development~\cite{Lyutikov:2011vc,Roy:2015kma,Inghirami:2016iru, Pu:2016bxy, Pu:2016ayh, Siddique:2019gqh, She:2019wdt}, and we leave this aspect in our future work.

The rest of this paper is organized as follows. In Sec.~\ref{sec:2} we present the (3+1)-dimensional ideal hydrodynamic model, the magnetic field profile, and the initial state of collisions. In Sec.~\ref{sec:3}, the numerical results on the elliptic flow $v_{2}$ at the LHC energies are presented. A brief summary is present in Sec.~\ref{sec:4}.

\section{Theory Framework}
\label{sec:2}

To simulate the QGP evolution, we employ the (3+1)-dimension viscous hydrodynamic model CLVisc~\cite{Pang:2018zzo,Wu:2021fjf}. For simplicity, no viscous corrections and net baryon current are included in the current study~\cite{Bali:2013owa,Pang:2016yuh}.

\subsection{Hydrodynamics}
\label{subsec:2-1}
In the presence of the squeezing force Eq.~(\ref{eq:F}), the hydrodynamic conservation equations can be written as~\cite{10.1063/1.478642}
\begin{equation}
\begin{aligned}
\partial_{\mu}T^{\mu\nu}=F^{\nu},
\label{eq:tmn}
\end{aligned}    
\end{equation}
where $T^{\mu\nu}$ is the energy momentum tensor. For ideal fluid, it reads
\begin{equation}
\begin{aligned}
T^{\mu\nu}=(\varepsilon+P) u^{\mu}u^{\nu}-Pg^{\mu\nu}, 
\label{eq:tmunu}
\end{aligned}    
\end{equation}
where $\varepsilon$ is the energy density, $P$ is the pressure, and $u^{\mu} = {\gamma} (1,\mathbf{v})$ represents the fluid velocity field. We would like to note that since we only interested in the effect of the interaction of the magnetic field with the QGP medium through the magnetization in the current study, we treat the magnetic field as an external field and do not consider the energy-momentum tensor of the magnetic field. Therefore, terms such as $\mathbf{B^2}$/2 and $B^{\mu}B^{\nu}$ are not included in the present work~\cite{Pang:2016yuh,Pu:2016ayh}. Hydrodynamic equations are solved using a lattice QCD-based Equation of State (EoS) provided by the Wuppertal-Budapest group~\cite{Borsanyi:2013bia}. As the system evolves below the switching temperature ($T_{\text{frz}}$ = 137~MeV), we use the Cooper-Frye formalism to obtain the hardon momentum distribution. The resonance decay contributions, as performed in Ref.~\cite{Pang:2018zzo}, are included in current work when calculating the yields and anisotropic flow for final charged hadrons. The space-time coordinate axes are defined as follow: proper time $\tau$, transverse plane coordinate $x$ (in the impact parameter direction), transverse plane coordinate $y$ (in direction perpendicular to the reaction plane), and longitudinal coordinate $\eta_\text{s}$ (the space-time rapidity). 

The relationship between magnetization $\mathbf{M}$ and magnetic field $\mathbf{B}$ can be expressed as $\mathbf{M} = \chi \mathbf{B}$. Lattice QCD calculations~\cite{Bali:2013owa,Bali:2020bcn} have shown that the magnetic susceptibility $\chi$ varies with both the magnetic field and temperature. Following the Refs.~\cite{Bali:2013owa}, the squeezing force density $F^{\nu}$ on the right-hand side of Eq.~(\ref{eq:tmn}) can be expressed as
\begin{equation}
\begin{aligned}
F^x=\frac{\chi}{2} {\partial_x}{\lvert{\mathbf{B}}\rvert}^2,~
F^y=\frac{\chi}{2} {\partial_y}{\lvert{\mathbf{B}}\rvert}^2,~
F^{\tau}=F^{\eta_\text{s}}=0.
\label{eq:Fxy}
\end{aligned}    
\end{equation}
Because the dynamics responsible for anisotropic flow are primarily determined by forces operating in the transverse plane, one may assume that the longitudinal force is negligible~\cite{Bali:2013owa}. This assumption aligns with the expectation that the magnetic field remains constant in the longitudinal direction within the QGP~\cite{Pang:2016yuh}. 

\begin{table}[!h]
\setlength{\tabcolsep}{1.5pt}
\begin{center}
\caption{\label{tab:parameter} The parameters of $\chi(T)$ in Eq.~(\ref{eq:chi})~\cite{Bali:2020bcn}}
\begin{tabular}{ |c|c|c|c|c|c|c|c|c| }
\hline
$\beta_1$  & $q_0$  & $g_0$    & $g_1$    &$g_2$  & $g_3$   & $g_4$  & $g_5$  & $h_3$\\ \hline
$1/(6{\pi}^2)$  & 0.1497  & 23.99   & -2.085  &0.1290  & 21.35  & -6.201   & 0.5766    & 0.1544  \\ 
\hline
\end{tabular}
\end{center}
\end{table}

The most recent lattice QCD results for the magnetic susceptibility $\chi$ for QCD matter can be found in Ref.~\cite{Bali:2020bcn}. The parameterization of $\chi$ that in agreement with perturbation theory at high temperatures reads 
\begin{equation}
\begin{aligned}
\chi(T)&= 2e^{2}\beta_1  \log \left(\frac{t}{q_0}\right) \\
 &~~~~\times \frac{1+g_0/t+g_1/t^2+g_2/t^3}{1+g_3/t+g_4/t^2+g_5/t^3} \exp(\frac{-h_3}{t}), 
\label{eq:chi}
\end{aligned}    
\end{equation}
where $t$ parameter is defined as $T$/(1GeV), and $T$ is the temperature. $e$ represents the elementary charge. 
It will be convenient to express the magnetic field in terms of $e$ because the combination $eB$ has units of GeV$^{2}$.
The parameters $\beta_1$, $q_0$, $g_i$ and $h_3$ are summarized in Table.~\ref{tab:parameter}.
\begin{figure}[htb!]
\includegraphics[width=6.5 cm]{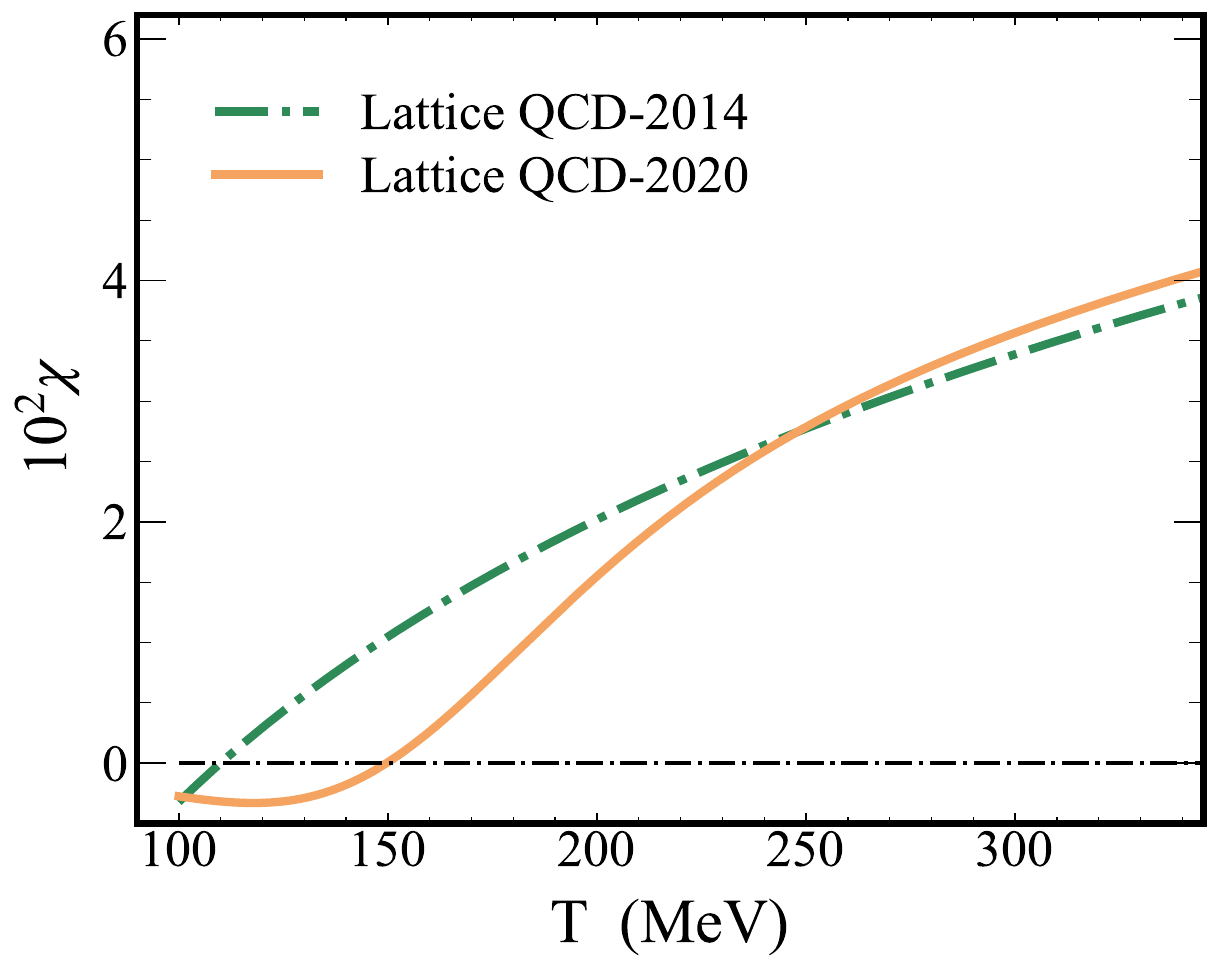}
\caption{(Color online) the magnetic susceptibility $\chi$ of QCD matter as a function of temperature. Results from lattice QCD calculation in 2014 (labeled as Lattice QCD-2014~\citep{Bali:2014kia,Borsanyi:2013bia}) and 2020 (labeled as Lattice QCD-2020~\citep{Bali:2020bcn}).}
\label{Fig:chi}
\end{figure} 

In Figure~\ref{Fig:chi},  we present the magnetic susceptibility $\chi$ as a function of temperature. The $\chi$ obtained from the lattice QCD calculations in 2014 and 2020 are presented for comparison. Notably, the parameterization of $\chi$ in 2020 reveals that below a temperature of 150 MeV, the QGP behaves as a diamagnetic material. To simplify our analysis, following previous work~\cite{Pang:2018zzo}, we apply a cutoff at $T=150$ MeV for $\chi$ to spec focus on the paramagnetic squeezing effect. 
The magnetic susceptibility $\chi$ in 2014 is described by $\chi(T) = \frac{e^2}{3{\pi}^2} \log{\frac{\text{T}}{\text{110~MeV}}}$, here T $\geq$ 110 MeV~\cite{Pang:2018zzo}. One can find that the $\chi(T)$ exhibits a nearly similar distribution for temperatures above 200 MeV.
This suggests a certain degree of consistency between the two datasets at high temperatures and indicates a similar impact on the anisotropic force density within the QGP under same external magnetic conditions (see below). However, further investigation is required to understand the effect of external magnetic on the QGP at temperatures below 150 MeV, which will be investigated in future work.

\subsection{Magnetic field profile}
\label{subsec:2-2}

To calculate the squeezing force density $F^{\nu}$, it is necessary to understand the spatial profile of the magnetic field at each point in time. 
However, the time evolution of the magnetic field is still subject to significant uncertainty~\cite{Pang:2016yuh}. 
Since the magnetic field generated by spectators drops rapidly in vacuum~\cite{Kharzeev:2007jp}, it has been suggested that the field could survive much longer in the presence of the QGP's non-zero electrical conductivity $\sigma_{el}$~\cite{Tuchin:2013ie,McLerran:2013hla,Li:2016tel}. This could be further prolonged by the charged quark-antiquark pairs formed via the gluon splitting and the Schwinger mechanism during the pre-equilibrium stage, see Refs.~\cite{Schwinger:1951nm,Tuchin:2013ie}. 



The medium's influence on the magnetic field's evolution is significant, but one can still explore its relationship with collision parameters using the Lienard-Wiechert potential for nucleon collisions in vacuum~\cite{Deng:2012pc}. Experimental results show weak correlation between the magnetic field $B$ and the electric charge number $Z$, but strong correlation with the impact parameter $b$ and the collision beam energy $\sqrt{s_\text{NN}}$~\cite{Bzdak:2011yy,Deng:2012pc,Kharzeev:2007jp,Li:2016tel}. People have also studied the magnetic field's spatial distribution within the reaction plane, finding anisotropic characteristics with rapid decrease along impact and slower change perpendicular to the plane~\cite{Deng:2012pc}.

\begin{table}[!h]
\begin{center}
\caption{\label{tab:Setup} The configurations for the space-time profile of the magnetic field.}
\begin{tabular}{l c c c c}
\hline\hline
Setup   & $\rho(\tau)$    & $\tau_\text{B}$~[fm]    & $\sigma_\text{x}$~[fm]   & $\sigma_\text{y}$~[fm]    \\ \hline
Setup-1~\citep{Pang:2016yuh}   & $\exp(-\frac{\tau}{\tau_\text{B}})$                  & 0.4         & 1.3         & 2.6      \\
Setup-2~\citep{Deng:2012pc}  & $\frac{1}{\left(1+\frac{{\tau}^2}{{\tau_\text{B}}^2}\right)}$   & 0.4         & 1.3         & 2.6   \\
Setup-3~\citep{Sun:2020wkg}   & $\frac{1}{\left(1+\frac{\tau}{\tau_\text{B}}\right)}$          & 0.4         & 1.3         & 2.6       \\
\hline\hline
\end{tabular}
\end{center}
\end{table}

Taking into account the above factors, the transverse distribution of the magnetic field, which exhibits longitudinal boost invariance, can be parametrized as~\citep{Deng:2012pc} 
\begin{equation}
\begin{aligned}
eB(x,y,\tau)=eB_0 \rho(\tau) \exp\left(-\frac{x^2}{2{\sigma_x}^2} -\frac{y^2}{2{\sigma_y}^2}\right) ,
\label{eq:eB}
\end{aligned}    
\end{equation}
where $eB_0$ represents the magnitude of the magnetic field at initial proper time $\tau_{0}$, it can be obtained as follow: 
\begin{equation}
\begin{aligned}
eB_0 = \frac{\gamma v_{z}Z\alpha}{R^{2}},
\label{eq:eB0}
\end{aligned}    
\end{equation}
here $\gamma$ is the Lorentz boost factor, $v_{z}$ is the beam velocity, $\alpha=1/137$ is the fine-structure constant, $Z$ is the electric charge number, and $R$ is the nuclear radius. For Pb nucleus, $Z=82$ and $R=6.62$ fm. Since the time evolution of the magnetic field $\rho(\tau)$ is still unknown, we adopt three decays widely used in previous studies~\cite{Deng:2012pc,Pang:2016yuh,Sun:2020wkg,Jiang:2022uoe,Sun:2023adv} in this paper. The Gaussian distribution widths along the $x$ and $y$ directions are set to $\sigma_{x}$ and $\sigma_{y}$~\citep{Deng:2012pc}. According to Refs.~\cite{Deng:2012pc,Pang:2016yuh}, we adopt $\sigma_{x}=1.3$ fm and $\sigma_{y}=2.6$ fm in centrality 20-50\% ($\left<b\right>\approx$ 10 fm) Pb+Pb collisions at the LHC, respectively.
We adopt three decay models from previous studies~\cite{Deng:2012pc,Pang:2016yuh,Sun:2020wkg,Jiang:2022uoe,Sun:2023adv} with a constant pre-equilibrium stage lifetime $\tau_B$ (also dubbed QGP formation time~\cite{Deng:2012pc}) to describe the magnetic field's time evolution. The parameters of magnetic field profile utilized in this paper are summarized in Table~\ref{tab:Setup}.

\begin{figure}[htb!]
\begin{center}
\includegraphics[height= 6 cm, width=7.3 cm]{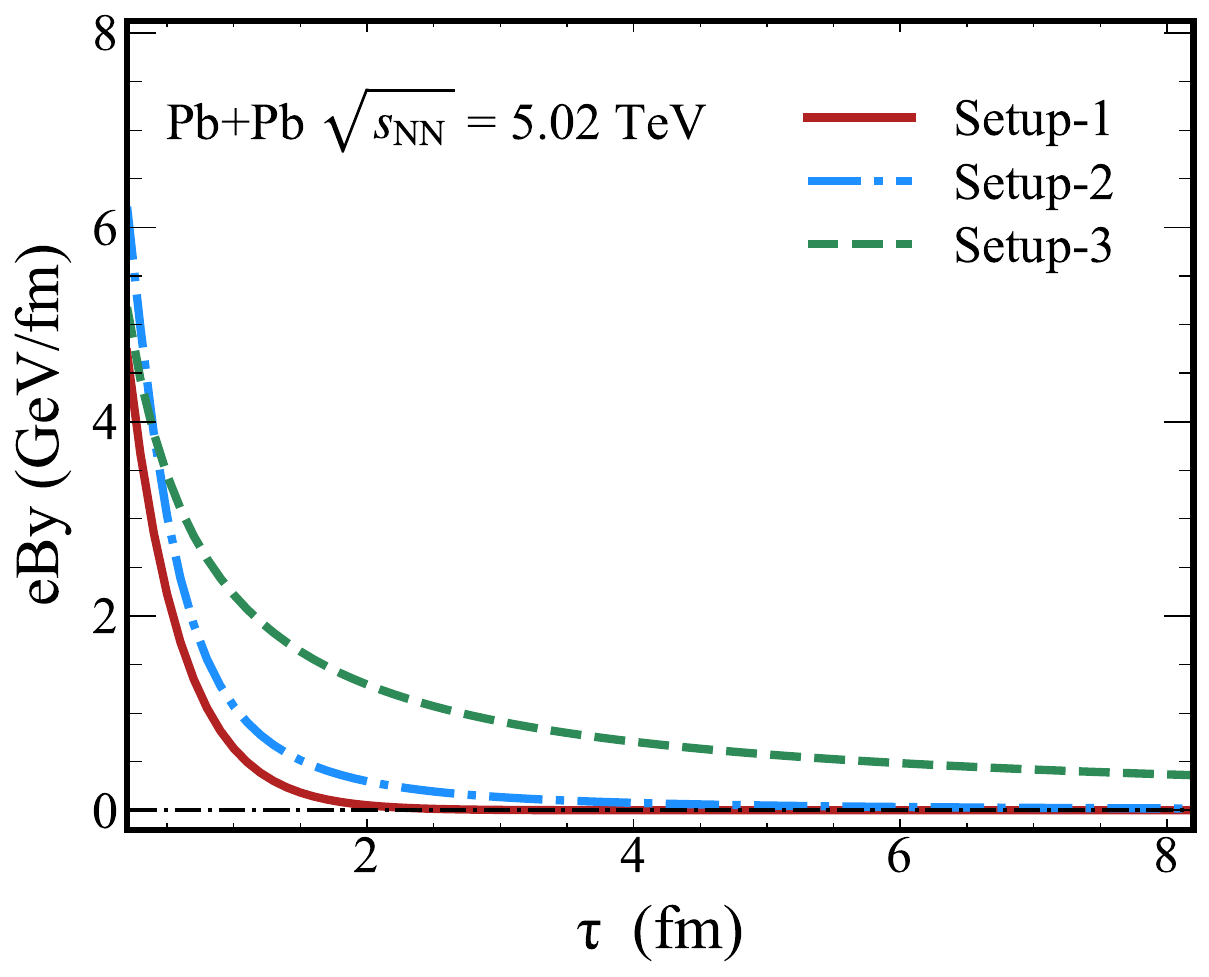}~\\
\includegraphics[height= 6 cm, width=7.5 cm]{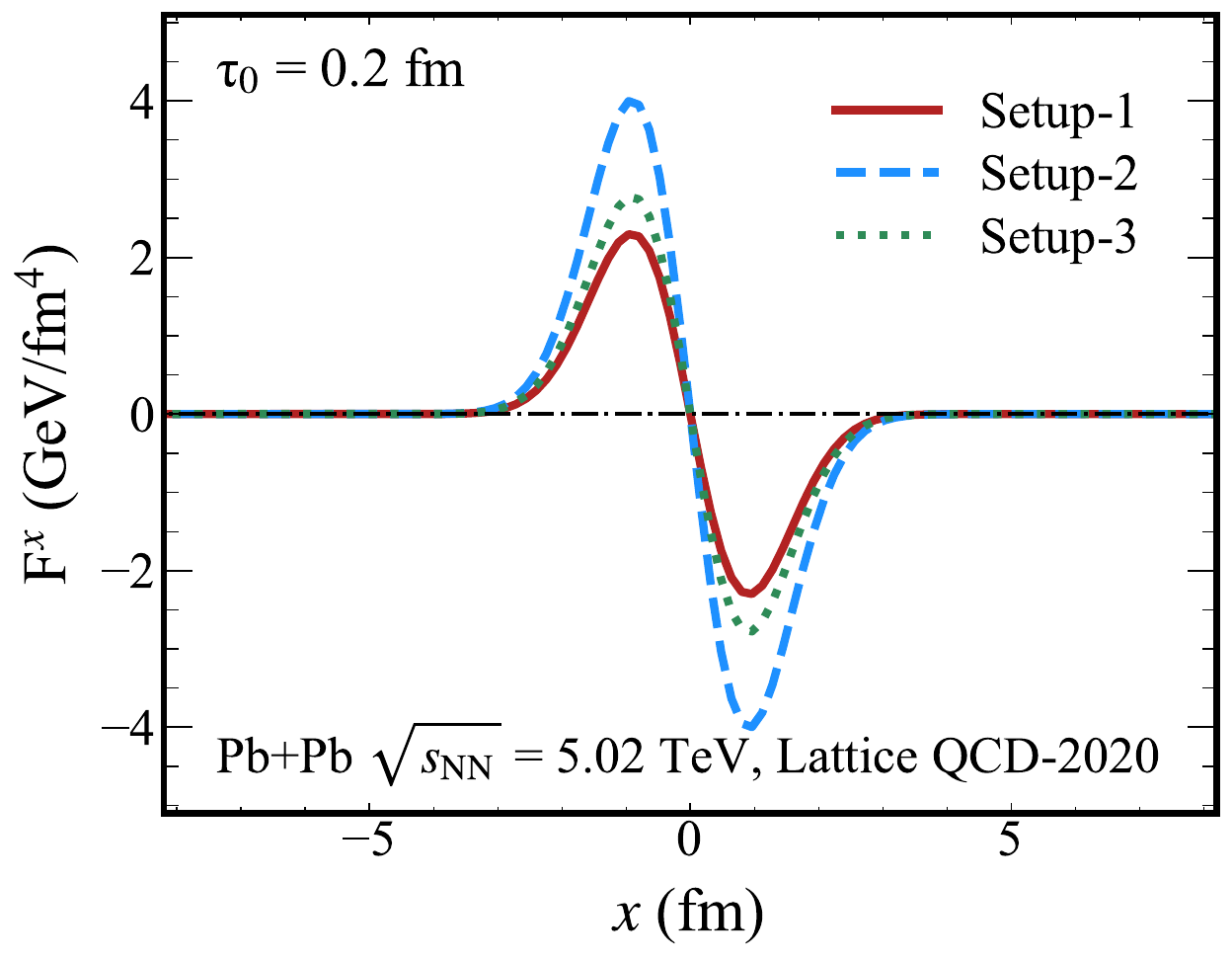}~\\
\includegraphics[height= 6 cm, width=7.5 cm]{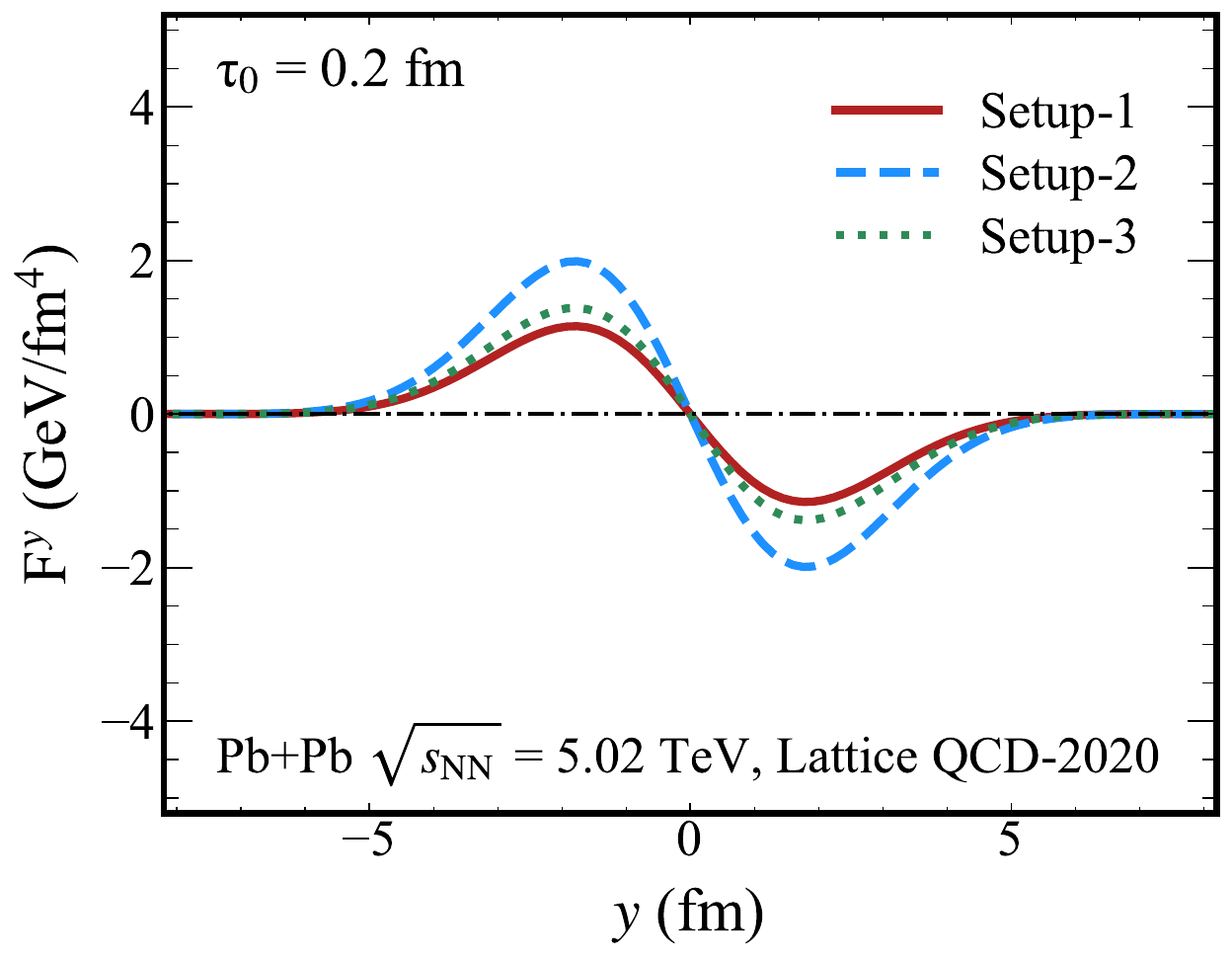}
\caption{(Color online) Upper panel: Time evolution of the magnetic field $eB_\text{y}$ in Pb+Pb collisions at $\sqrt{s_\text{NN}} = 5.02$~TeV ($b=$ 10 fm) under magnetic field Setup-1, 2, and 3. Middle and lower panel: Squeezing force density $F^{x}$ ($F^{y}$) along the $x$-axis ($y$-axis) in Pb+Pb collisions at $\sqrt{s_\text{NN}} = 5.02$~TeV with $\tau_\text{0} = 0.2$~fm and $b=$ 10 fm, obtained from Lattice QCD-2020 under magnetic field Setup-1, 2, and 3.\label{Fig:magnetic}}
\end{center}
\end{figure}

\begin{figure*}[!htb]
\includegraphics[width=8.0 cm]{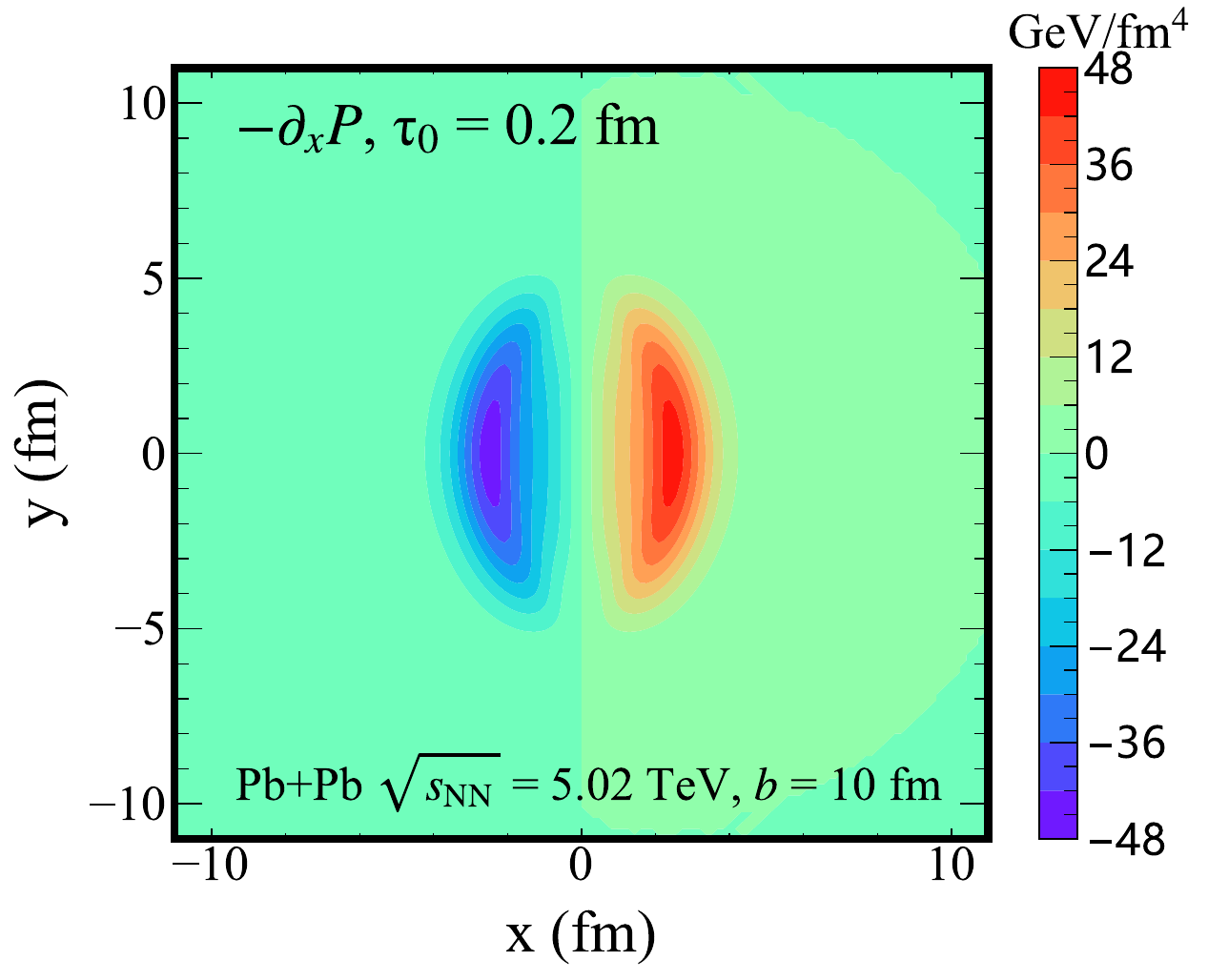}~
\includegraphics[width=8.0 cm]{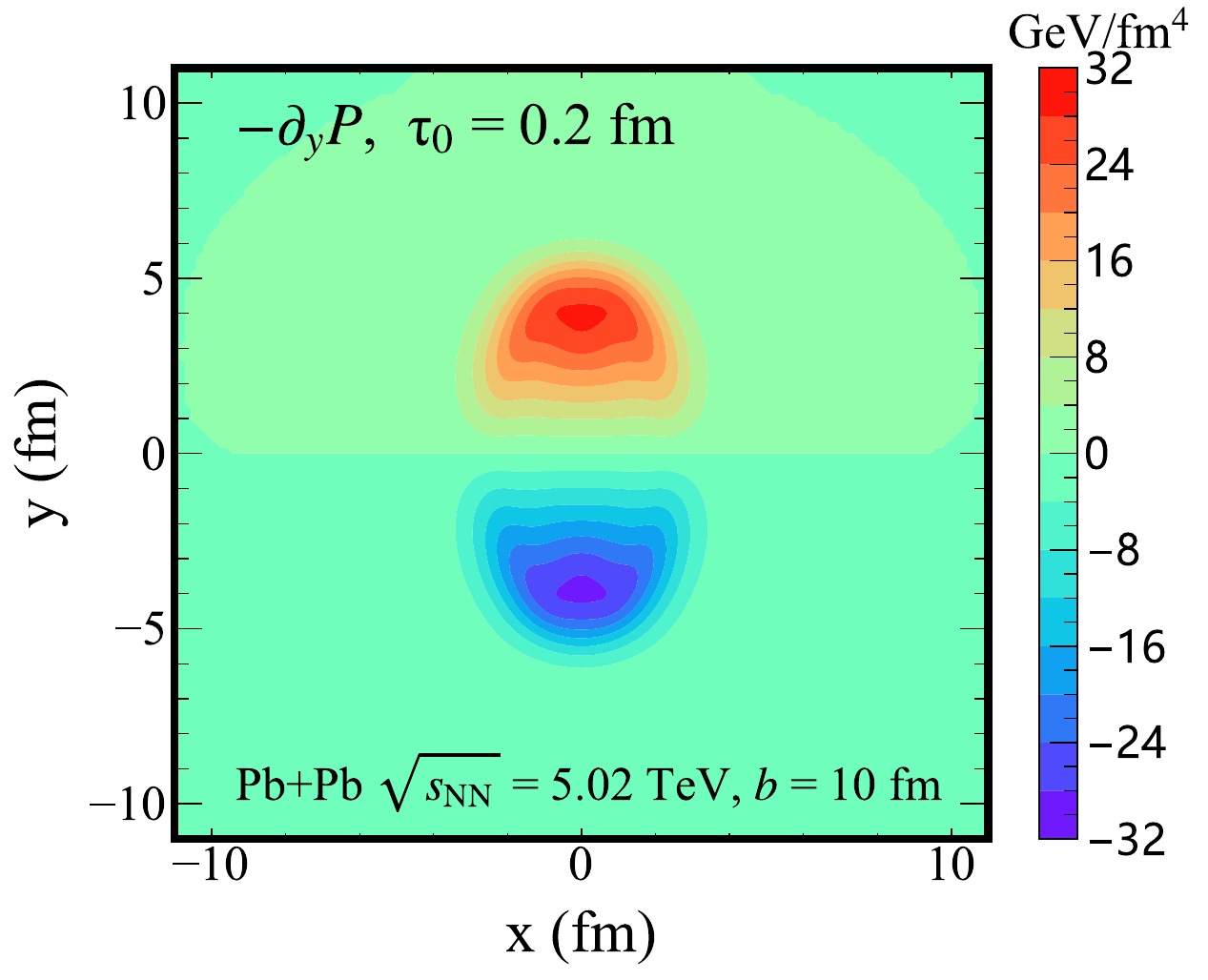} \\
\includegraphics[width=8.0 cm]{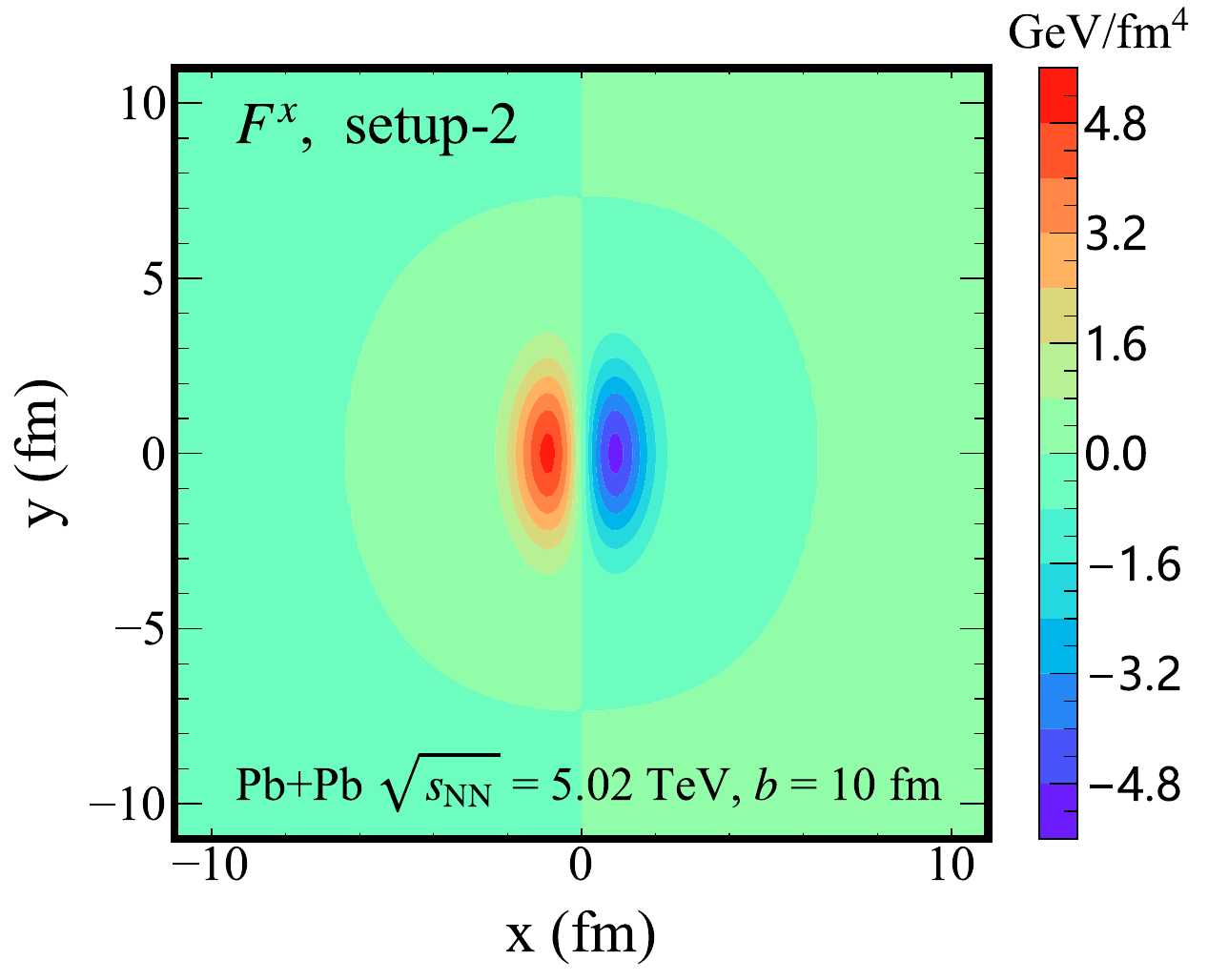}~
\includegraphics[width=8.0 cm]{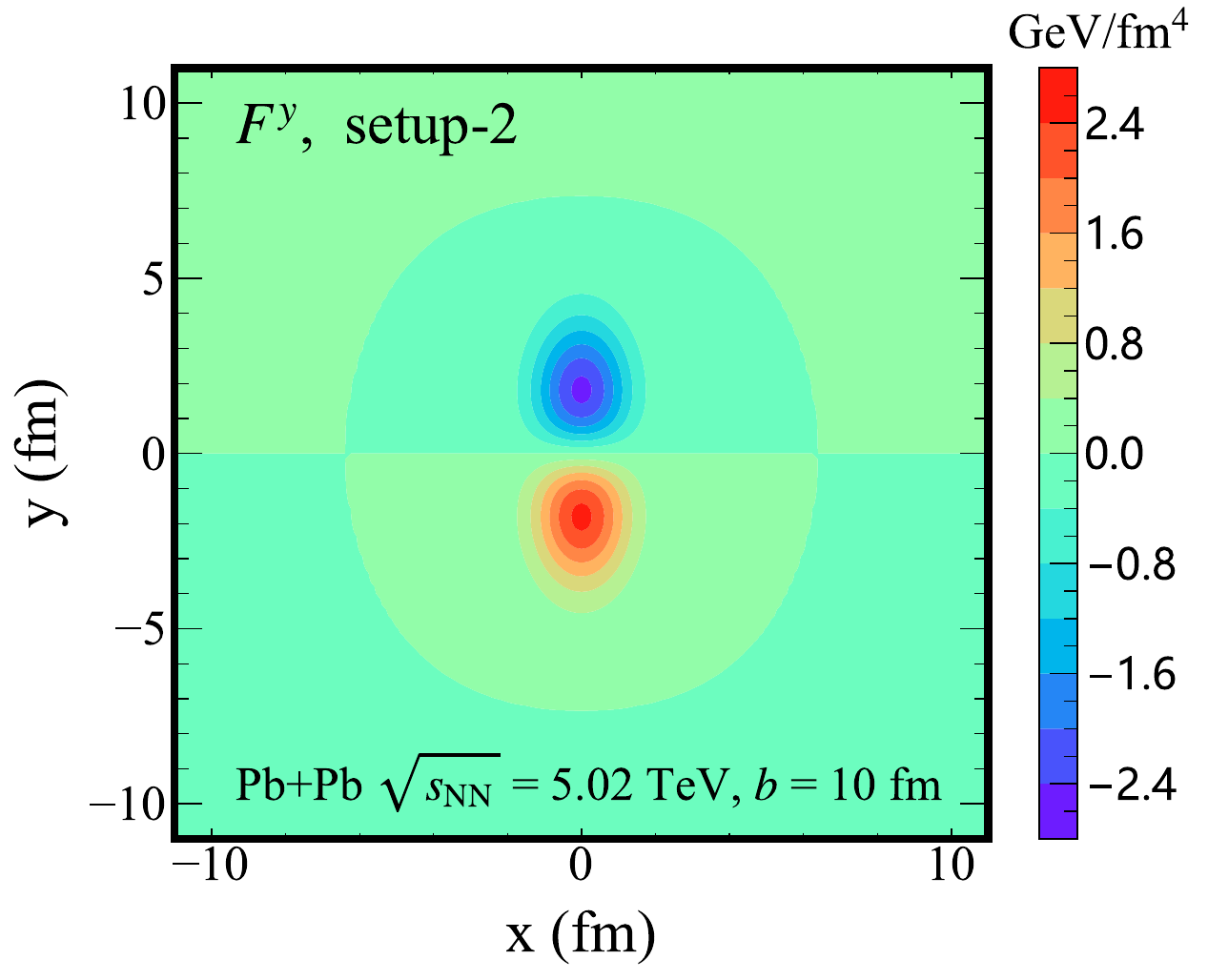}
\caption{(Color online) The pressure gradients (top left) along the $x$ direction and (top right) along the $y$ direction, as well as the squeezing force density (bottom left) along the $x$ direction and (bottom right) along the $y$ direction for Pb+Pb collisions at $\sqrt{s_\text{NN}}= 5.02$~TeV with impact parameter $b$ = 10~fm under magnetic field setup-2.\label{Fig:Fpxy}}
\end{figure*} 

In the upper panel of Figure~\ref{Fig:magnetic}, we present the time evolution of magnetic fields in centrality 20-50\% Pb+Pb collisions at $\sqrt{s_\text{NN}}=5.02$~TeV, assuming three magnetic field configurations in Table~\ref{tab:Setup}.
One may clearly observe that $eB_{y}$ exhibits the slowest decay in Setup-3, followed by Setup-2, while it decays the most rapidly in Setup-1.

In the middle panel of Figure~\ref{Fig:magnetic}, we present the squeezing force density $F^{x}$ at $\tau_{0}=0.2$~fm, utilizing the $\chi$ values obtained from lattice QCD calculations conducted in 2020. One finds that the magnetic field Setup-2 exhibits the highest value, followed by Setup-3, with Setup-1 having the lowest. The same analysis can be applied to $F^{y}$ (lower panel).

\subsection{Initial state}
\label{subsect2-3}

Using the optical Glauber model, we construct the initial energy density distribution as follows:
\begin{equation}
\begin{aligned}
\varepsilon(\tau_\text{0},x,y,\eta_\text{s})= K(0.95N_{{wn}}+0.05N_{bc})H(\eta_\text{s}) 
\label{eq:varepsilon}
\end{aligned} 
\end{equation}
where $K$ is a overall factor which constrained by the multiplicity distribution ($dN_{\textrm{ch}}/d\eta$ or $dN_{\textrm{ch}}/dy$), $N_{wn}$ represents the number of wounded nucleons, $N_{bc}$ represents the number of binary collisions~\cite{Pang:2016yuh,Jiang:2023vxp}. 
Following Refs.~\cite{Pang:2016yuh,Jiang:2021foj}, we use a empirical $H(\eta_\text{s})$ function to describe the plateau structure of $dN/d\eta$ for charged hadrons in the most central region.
This function reads 
\begin{equation}
\begin{aligned}
H(\eta_\text{s})=\exp\left[-\frac{(\left|\eta_\text{s} \right|-\eta_\omega)^2}{2\sigma_\eta^2}\theta(\left|\eta_\text{s}\right|-\eta_\omega)\right],
\label{eq:8}
\end{aligned}
\end{equation}
where $\eta_\omega=2.2$ is the width of the platform at the middle rapidity, and $\sigma_{\eta}=1.8$ is the width of the Gaussian decay outside the platform region~\cite{Jiang:2021foj}. We assume $\eta_\omega$ and $\sigma_{\eta}$ remain the same across different collision systems.
In calculating $N_{wn}$ and $N_{bc}$, the inelastic scattering cross section $\sigma_\text{0}$ is set to 62~mb, 67mb and 75mb for Pb+Pb at 2.76 TeV, 5.02~TeV and 10.6 TeV collisions~\cite{Loizides:2017ack}.  

In the optical Glauber model, the nucleon density of the Pb nucleus is described by the Woods-Saxon distribution, 
\begin{equation}
\begin{aligned}
\rho(r)=\frac{n_\text{0}}{\exp(\frac{r-R}{d})+1},
\label{eq:rho}
\end{aligned}    
\end{equation}
where $n_{0}=0.17$ fm$^{-3}$ denotes the average nucleon density, $r=\sqrt{x^{2}+y^{2}+z^{2}}$ is the radial position with $x,~y,~z$ being the space coordinates, $R$ is the nuclear radius, and $d=0.546$ fm stands for the diffusion rate for Pb nucleus. 

\begin{table}[!h]
\begin{center}
\caption{\label{tab:energy} The $K$ parameter for the most central Pb+Pb collisions at the LHC}
\begin{tabular}{|l| c| c |c| }
\hline
$~~~~\snn$        & 2.76 TeV     & 5.02 TeV    & 10.6 TeV        \\ \hline
$K$ [CeV/fm$^3$]   & 490.1    & 590.1   & 765.6    \\
\hline
\end{tabular}
\end{center}
\end{table}

In Table.~\ref{tab:energy}, we have listed the value of $K$ parameter at the initial thermalization time $\tau_\text{0}=0.2$ fm. The value of $K$ parameters for $\snn=2.76$ TeV and $\snn=5.02$ TeV are constrained by the experimental multiplicity distribution ($dN_{\textrm{ch}}/d\eta$) at the most central collisions. To calibrate the $K$ parameter for $\snn=10.6$ TeV at the HL-LHC energy region, we use the empirical formula from ALICE Collabration~\cite{ALICE:2022wpn}, 
\begin{equation}
\begin{aligned}
\frac{2}{\left<N_{\text{part}}\right>} \left.\frac{dN_{\text{ch}}}{d\eta} \right|_{|\eta_{s}|<0.5} = A(\snn)^{2p}/2, 
\label{eq:k}
\end{aligned}    
\end{equation}
for the most central 0-5\% Pb+Pb collisions. Here $\left<N_{\text{part}}\right>$ is the averaged number of participant nucleons~\cite{Loizides:2017ack}. The $\snn$ is the collision energy of a pair of nucleons in their center-of-mass frame. 
The parameter $A=0.745$ and $p=0.152$ are utilized in this study~\cite{Citron:2018lsq,ALICE:2018cpu}. For Pb+Pb collisions at $\snn=10.6$ TeV, the $dN_{\text{ch}}/{d\eta}$ at the mid-rapidity region $|\eta_{s}|<0.5$ is approximately 2408, which is consistent with the CERN Yellow Report result~\cite{Citron:2018lsq}. Specifically, we assume the impact parameter $b$ = 2.6~fm for centrality 0-5\% Pb+Pb collisions in this work~\cite{Loizides:2017ack}. 

With above initial state settings, Figure~\ref{Fig:Fpxy} shows a comparison of pressure gradients and magnetic-induced squeezing force density at initial thermalization time $\tau_{0}=0.2$ fm.
The initial pressure gradients ($-\partial_{x}P$ and $-\partial_{y}P$) in the transverse plane for Pb+Pb collisions at $\sqrt{s_\text{NN}}=5.02$~TeV with impact parameter $b=10$ fm are shown in Figure~\ref{Fig:Fpxy} upper panel. 
One may clearly find that the pressure gradient is higher along the $x$ direction than along the $y$ direction, peaking at approximately 48 GeV/fm$^{4}$.

In the lower panel of Figure~\ref{Fig:Fpxy}, we present the squeezing force densities ($F^{x}$ and $F^{y}$) in the transverse plane for Pb+Pb collisions at $\snn=5.02$ TeV ($b=10$ fm). Here we employed magnetic field Setup-2, which provides the highest initial magnetic field at $\tau_{0}=0.2$ fm. The squeezing force density was computed using the local temperature obtained from hydrodynamic simulations, based on Eq.~(\ref{eq:chi}) (Lattice QCD-2020). We find that the squeezing force densities and pressure gradients have similar shapes, with the former accounting for about 10\% of the latter. 


Through a comparison of the pressure gradient (upper) and squeezing force density (lower) in Figure~\ref{Fig:Fpxy}, it is evident that, as expected, the squeezing force density and pressure gradient along the $x$ direction are greater than those along the $y$ direction. The maximum values of these quantities are approximately 5.6 $\mathrm{GeV/{fm}^4}$ and 48$\mathrm{GeV/{fm}^4}$ along the $x$ direction, respectively. A further comparison of the top and bottom plane in Figure~\ref{Fig:Fpxy} reveals that the squeezing force densities exhibit an opposite trend to that of the pressure gradients in both the $x$ and $y$ directions, thus leading to a paramagnetic squeezing effect at the initial stage. 

Moreover, in magnetic field setup-2 (as detailed in Table.~\ref{tab:Setup}), where $2\sigma_{x} = \sigma_{y} = 2.6$~fm, Figure~\ref{Fig:Fpxy} shows that the maximum squeezing force density is observed at $x = \pm1$~fm, whereas the maximum pressure gradient is located at $x = \pm~2.5$~fm. Notably, the region of high pressure gradient extends further than the region where the squeezing force density is significant. Consequently, it is necessary to adjust the values of $\sigma_x$ and $\sigma_y$ to modify the squeezing force density distribution (see Subsection~\ref{subsec:3-4}) and study its impact on elliptic flow $v_{2}$.

\subsection{Anisotropy flow}
\label{subsec:2-4}

In non-central collisions, the pressure gradients cause faster expansion of QGP along $x$ than $y$ direction, leading to higher transverse momentum of final charged hadrons. 
The squeezing force density from external magnetic fields also affects this anisotropic expansion. Therefore, the elliptic flow $v_2$ as a function of transverse momentum of final charged hadrons is a key observable to study the momentum anisotropy and squeezing force density~\citep{Bali:2014kia,Borsanyi:2013bia,Pang:2016yuh}.
  
The $p_\text{T}$ differential elliptic flow $v_2$ of final charged hadrons, which is defined as, 
\begin{equation}
\begin{aligned}
v_2(p_T) \equiv \frac{\int d\phi \frac{dN}{dY dp_T d\phi}\cos\left[2(\phi-\mathbf{\Psi}_2)\right]}{\int d\phi \frac{dN}{dY dp_T d\phi}}.
\label{eq:v2}
\end{aligned}   
\end{equation}
Here, since we utilize a smooth initial condition for generating the initial energy density in QGP simulations, 
resulting in $\mathbf{\Psi}_2$ equals zero and minimal event-by-event fluctuations.
In the following Sec.~\ref{sec:3}, we will discuss in detail the dependence of the squeezing effect reveal by $v_2$ on different magnetic field setups.

\section{Numerical Results}
\label{sec:3}


\begin{figure}[htb!]
\includegraphics[height=6 cm, width=8 cm]{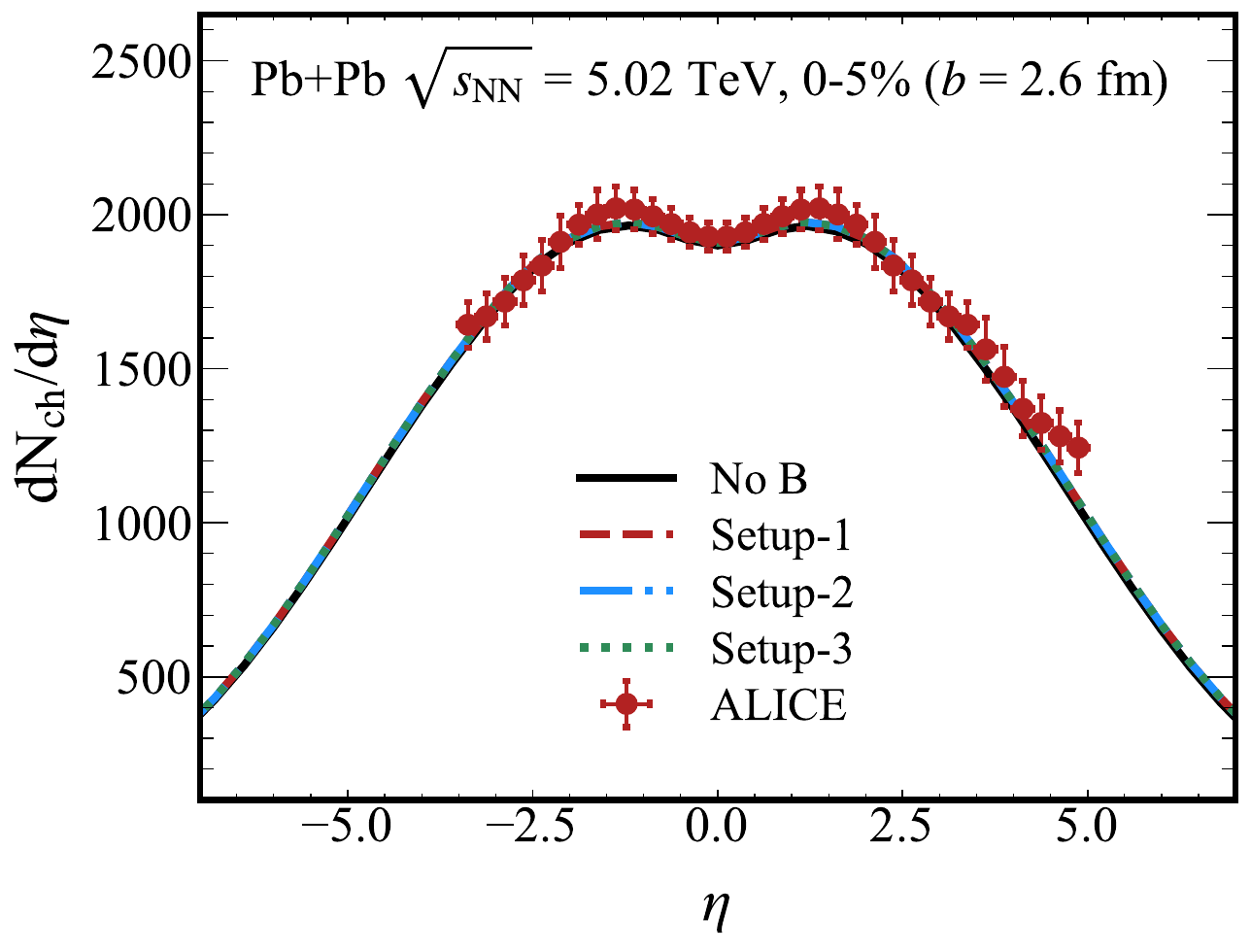}~\\
\includegraphics[height=6 cm, width=8 cm]{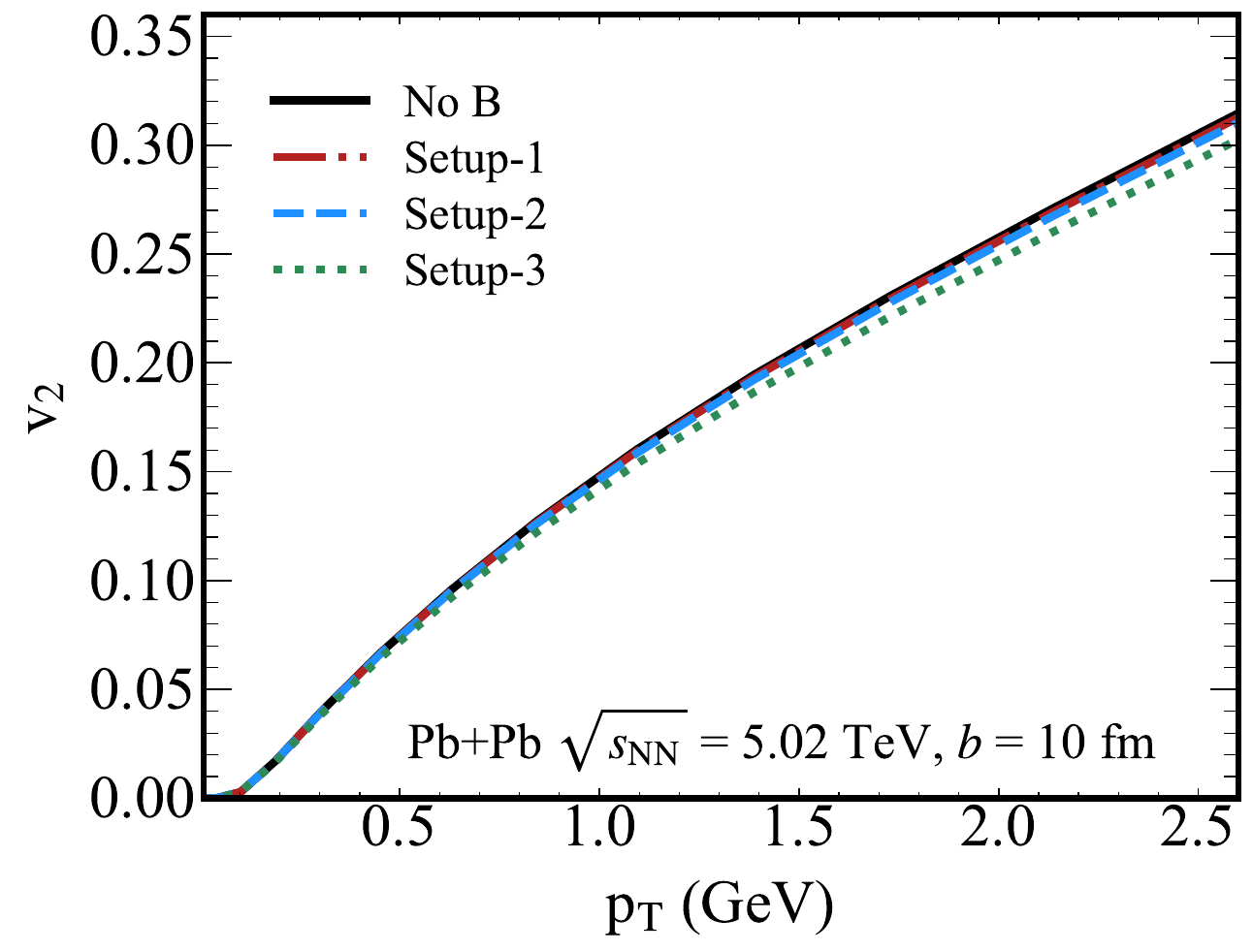}
\caption{(Color online) Upper panel: Pseudorapidity distribution $dN_{ch}/d\eta$ of final charged hadrons in $\sqrt{s_\text{NN}}=5.02$~TeV Pb+Pb collisions at centrality 0-5\% (b=2.6~fm) under no magnetic field (No B), magnetic field Setup-1, Setup-2, Setup-3 and ALICE Collaboration~\cite{ALICE:2016fbt}. Lower panel: Elliptic flow $v_2$ of final charged hadrons as a function of transverse momentum $p_\text{T}$ in $\sqrt{s_\text{NN}}=5.02$~TeV Pb+Pb collisions (b=10~fm) with no magnetic field (No B), magnetic field Setup-1, Setup-2, and Setup-3.\label{Fig:eta-v2}}
\end{figure} 

\subsection{The impact of magnetic field lifetime}
\label{subsec:3-1}
We focus on the squeezing effect of different magnetic field on the final charged hadron in Pb+Pb collisions at $\snn=$ 5.02 TeV in this section.

In the upper panel of Figure~\ref{Fig:eta-v2}, we show the pseudorapidity distribution of final charged hadrons at centrality 0-5\% ($b=2.6~$fm) in $\sqrt{s_\text{NN}}=5.02$~TeV Pb+Pb collisions. For calibration, we have utilized the experimentally measured pseudo-rapidity distribution $dN_{ch}/d\eta$ by the ALICE collaboration. Notably, all calculations with different magnetic field setups (No B, Setup-1, Setup-2, and Setup-3) produce the same charged multiplicity in the most central collisions.

Since the spatial distribution of squeezing force density and pressure gradient, which vary with collision centrality, suggest that previous comparisons of multiplicity density $dN_{ch}/d\eta$ at the most central collisions do not fully capture their impact. To account for the interplay between squeezing force density and pressure gradient, we calculate the elliptic flow $v_2$ of final charged hadrons in 20-50\% centrality ($b=10$~fm) Pb+Pb collisions.

In the lower panel of Figure~\ref{Fig:eta-v2}, we present the elliptic flow $v_2$ of final charged hadrons as a function of transverse momentum $p_\text{T}$ under different magnetic field setups: No B (without magnetic field), Setup-1, Setup-2, and Setup-3. The squeezing force density with magnetic field Setup-2 generates a 2\% suppression effect, while Setup-3 has a greater suppression effect on the elliptic flow $v_2$, reducing it by approximately 3-4\%. This is because Setup-3 has the longest lifetime (as shown in Figure~\ref{Fig:magnetic}). We also note that the $\tau_{B}=0.4$ fm used in Setup-1 is smaller than that employed in Ref.~\citep{Pang:2016yuh}, leading to a smaller suppression of $v_{2}$ in this study for Setup-1. When comparing the $v_{2}$ results for No B, Setup-1, Setup-2, and Setup-3, we find that the shorter the lifetime of the magnetic field, the smaller the effect of squeezing force density on $v_{2}$ suppression. Since Setup-2 has been commonly utilized in previous study~\cite{Deng:2012pc}, we will primarily concentrate on the squeezing force density induced by magnetic field Setup-2 in the subsequent sections.

\subsection{The impact of magnetic susceptibility}
\label{subsec:3-2}
\begin{figure}[htb!]
\includegraphics[height=6 cm, width=8 cm]{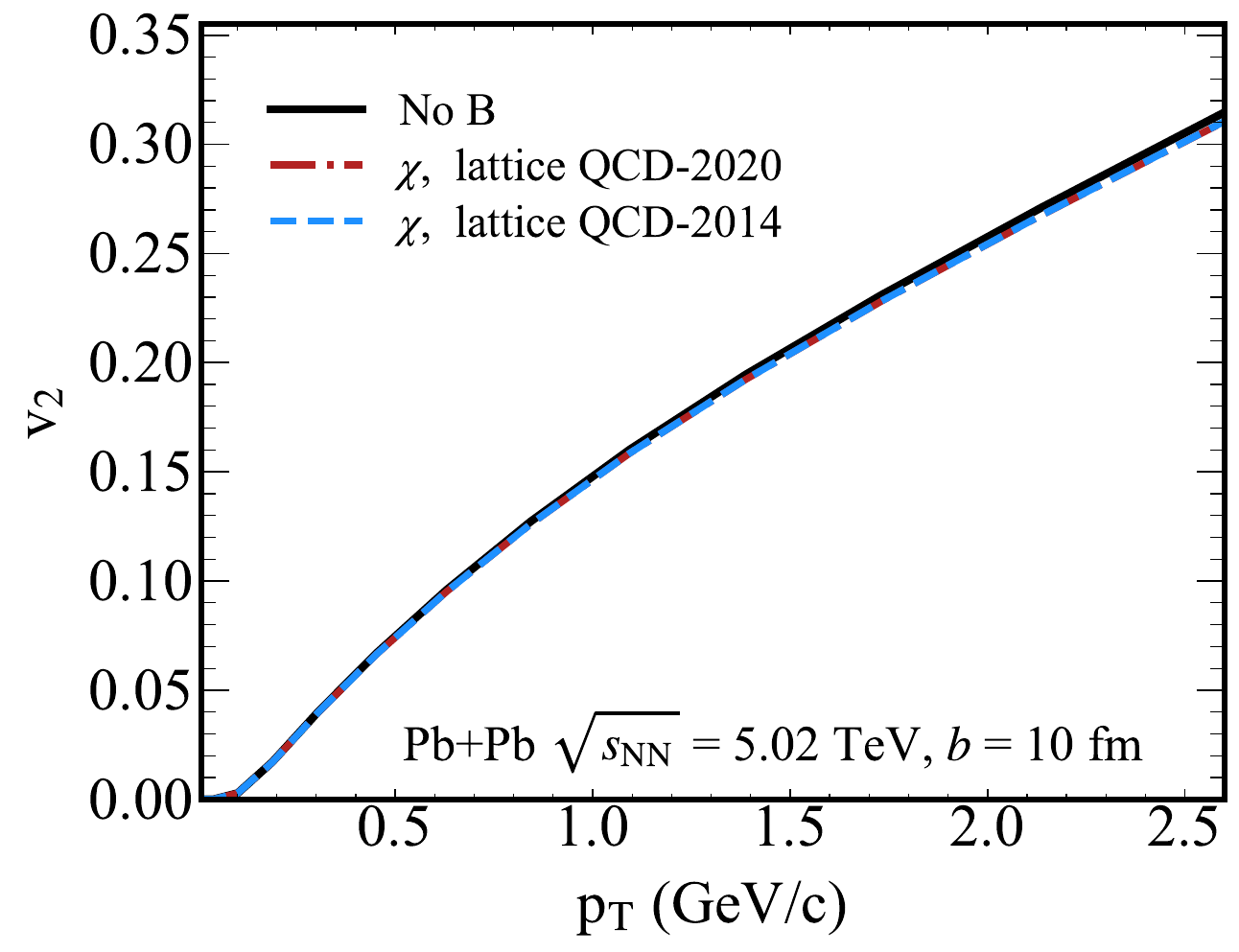}
\caption{(Color online) The elliptic flow $v_2$ as a function of $p_\text{T}$ for $\sqrt{s_\text{NN}}=5.02$~TeV Pb+Pb collisions at centrality 20-50\% (b=10~fm) under no magnetic field (No B) and under magnetic field Setup-2 with magnetic susceptibility $\chi$ obtained from lattice QCD-2014 and lattice QCD-2020 results.\label{Fig:v2-chi}}
\end{figure}  

In this section, we explore how magnetic susceptibility affects the squeezing effect and the elliptic flow.

In Figure~\ref{Fig:v2-chi}, we present the elliptic flow $v_2$ as a function of transverse momentum $p_\text{T}$ with magnetic susceptibility $\chi$ in lattice QCD-2014 and lattice QCD-2020 for $\sqrt{s_\text{NN}}=5.02$~TeV Pb+Pb collisions under magnetic field Setup-2, and compared it with the case without magnetic field (No B).
One can observe that the squeezing force density, when applied to both $\chi$ lattice QCD-2020 and $\chi$ lattice-QCD 2020 under magnetic field Setup-2, results in a minor reduction of the elliptic flow $v_{2}$ by only 1-2\%. 
This is due to the fact that the magnetic susceptibility $\chi$ exhibits minimal variation in the high temperature region (as seen in Figure~\ref{Fig:chi}). In contrast, in the low temperature region, the magnetic field is almost non-existent, being both very weak and close to zero.

\subsection{The impact of magnetic field spatial distribution}
\label{subsec:3-4}
\begin{figure}[htb!]
\includegraphics[height=6 cm, width=8 cm]{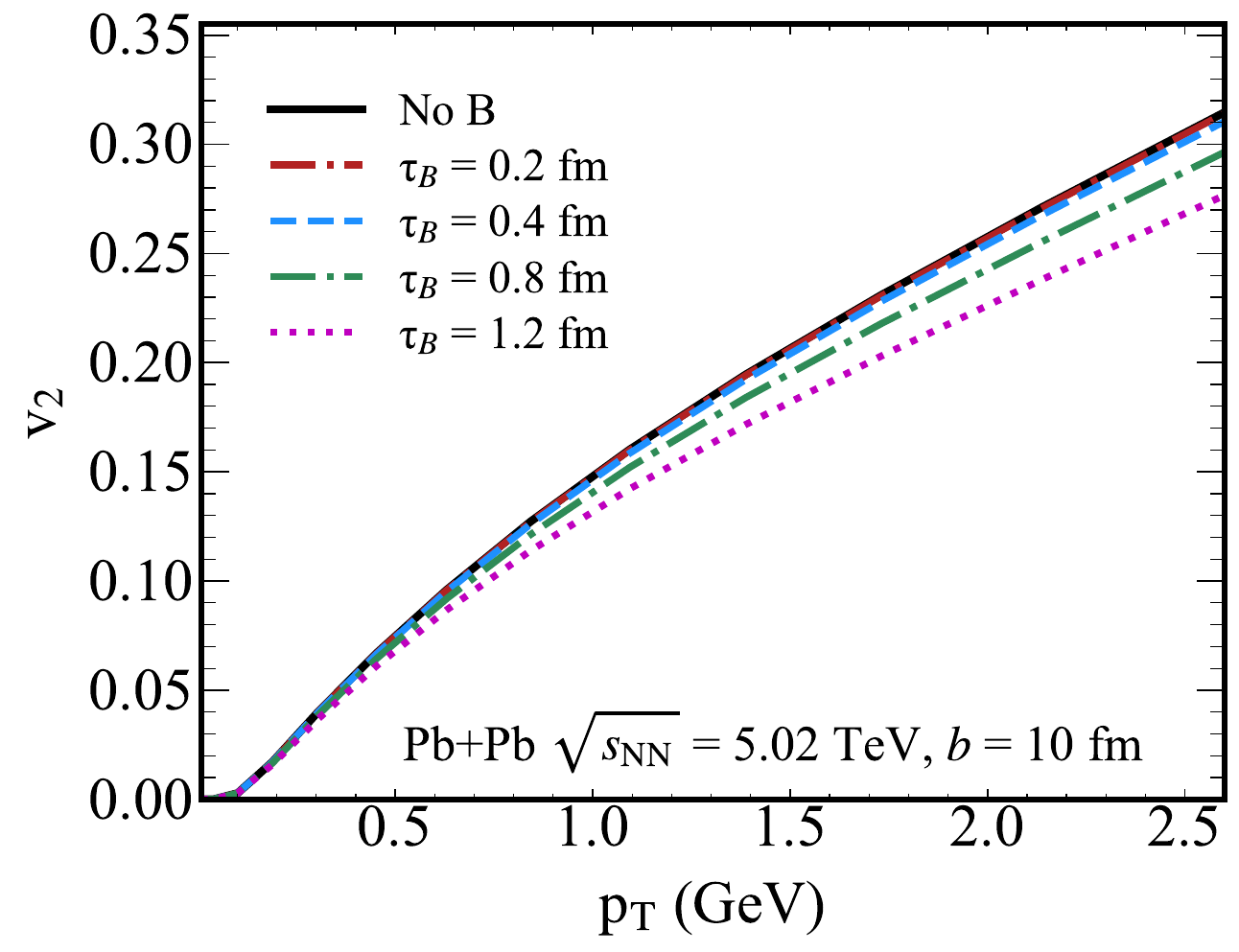}~~\\
\includegraphics[height=6 cm, width=8 cm]{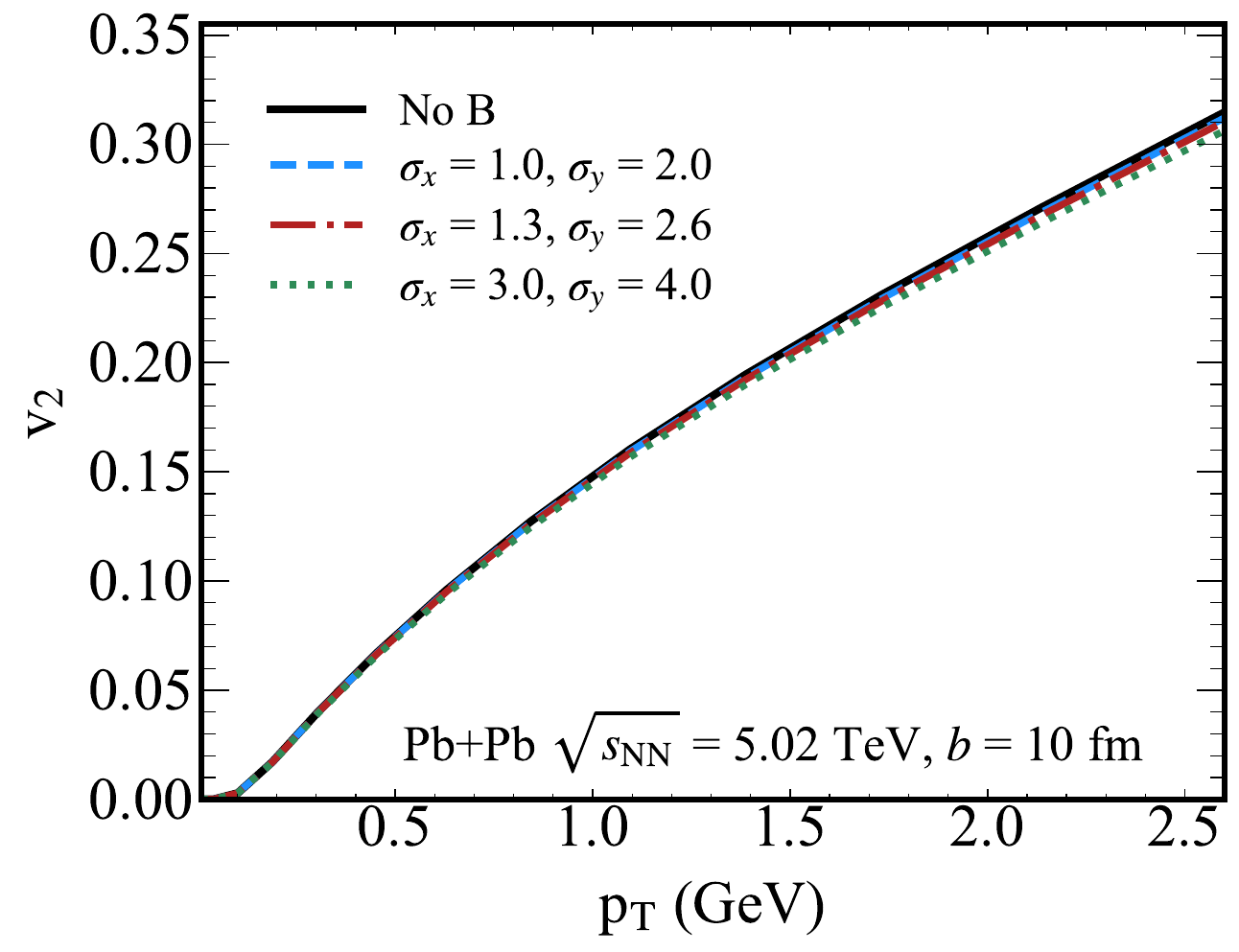}
\caption{(Color online) Upper panel: Elliptic flow $v_2$ of final charged hadrons as a function of transverse momentum $p_\text{T}$ in $\sqrt{s_\text{NN}}=5.02$~TeV Pb+Pb collisions (b=10~fm) with no magnetic field (No B) and different QGP formation time $\tau_{B}$ under magnetic field setup-2. 
Lower panel:Elliptic flow $v_2$ of final charged hadrons as a function of transverse momentum $p_\text{T}$ in $\sqrt{s_\text{NN}}=5.02$~TeV Pb+Pb collisions (b=10~fm) with no magnetic field (No B) and different Gaussian distribution widths along the $x$ and $y$ directions under magnetic field Setup-2. \label{Fig:taub-sig}}
\end{figure}  

We further investigate the impact of magnetic field configuration on the squeezing effect under Setup-2 and its influence on the elliptic flow $v_{2}$ of charged hadrons in $\sqrt{s_\text{NN}}= 5.02$~TeV Pb+Pb collisions ($b=10$~fm).

In the upper panel of Figure~\ref{Fig:taub-sig}, we show the elliptic flow $v_2$ as a function of the transverse momentum $p_\text{T}$  for various QGP formation times. One can clearly observe that as the QGP formation time $\tau_\text{B}$ increases from 0.4 fm to 1.2 fm, the elliptic flow $v_2$ decreases. When $\tau_\text{B}=1.2$~fm, the suppression of $v_{2}$ becomes around 10\%. This trend can be attributed to the fact that a longer $\tau_\text{B}$ extends the lifetime of the external magnetic field, resulting in a longer squeezing force density for the QGP fireball. Consequently, this stronger squeezing effect suppressed the elliptic flow $v_2$. 

In the lower panel of Figure~\ref{Fig:taub-sig}, we present the elliptic flow $v_2$ versus transverse momentum $p_\text{T}$ for different Gaussian distribution widths along the $x$ and $y$ directions. 
One finds that as the Gaussian distribution widths $\sigma_x$ and $\sigma_y$ increases, the spatial size of the magnetized region increases, then suppressed the elliptic flow $v_2$ about 3\%. 

This suggests that either prolonging the lifetime of the external magnetic field (i.e., increasing $\tau_{B}$) or extending its spatial size (i.e., increaing $\sigma_{x}$ and $\sigma_{y}$) would result in a stronger squeezing force density, ultimately leading to the suppression of $v_{2}$. We also emphasize that the profile of the electromagnetic field in heavy ion collisions is a complex and ongoing question~\cite{Huang:2022qdn}. Therefore, the analysis presented here may serve as a simplified case for exploration. Nevertheless, if one employs a different magnetic field configuration from various theoretical frameworks, the general behavior of its impact on the squeezing effect is expected to remain consistent with this work.

\subsection{The squeezing effect in different center-of-mass energy}
\label{subsec:3-5}
\begin{figure}[htb!]
\includegraphics[height=6 cm, width=8 cm]{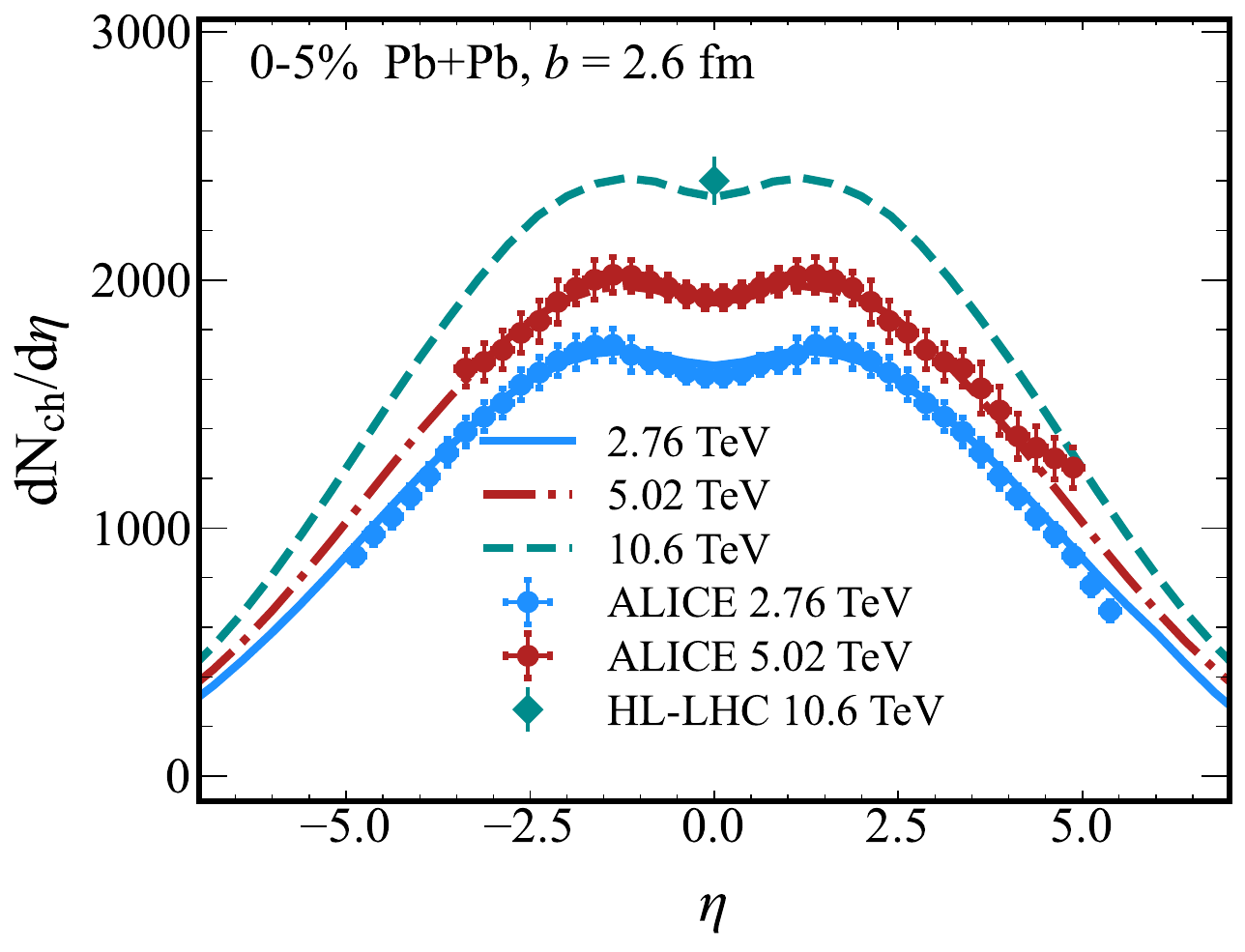}~\\
\includegraphics[height=6 cm, width=8 cm]{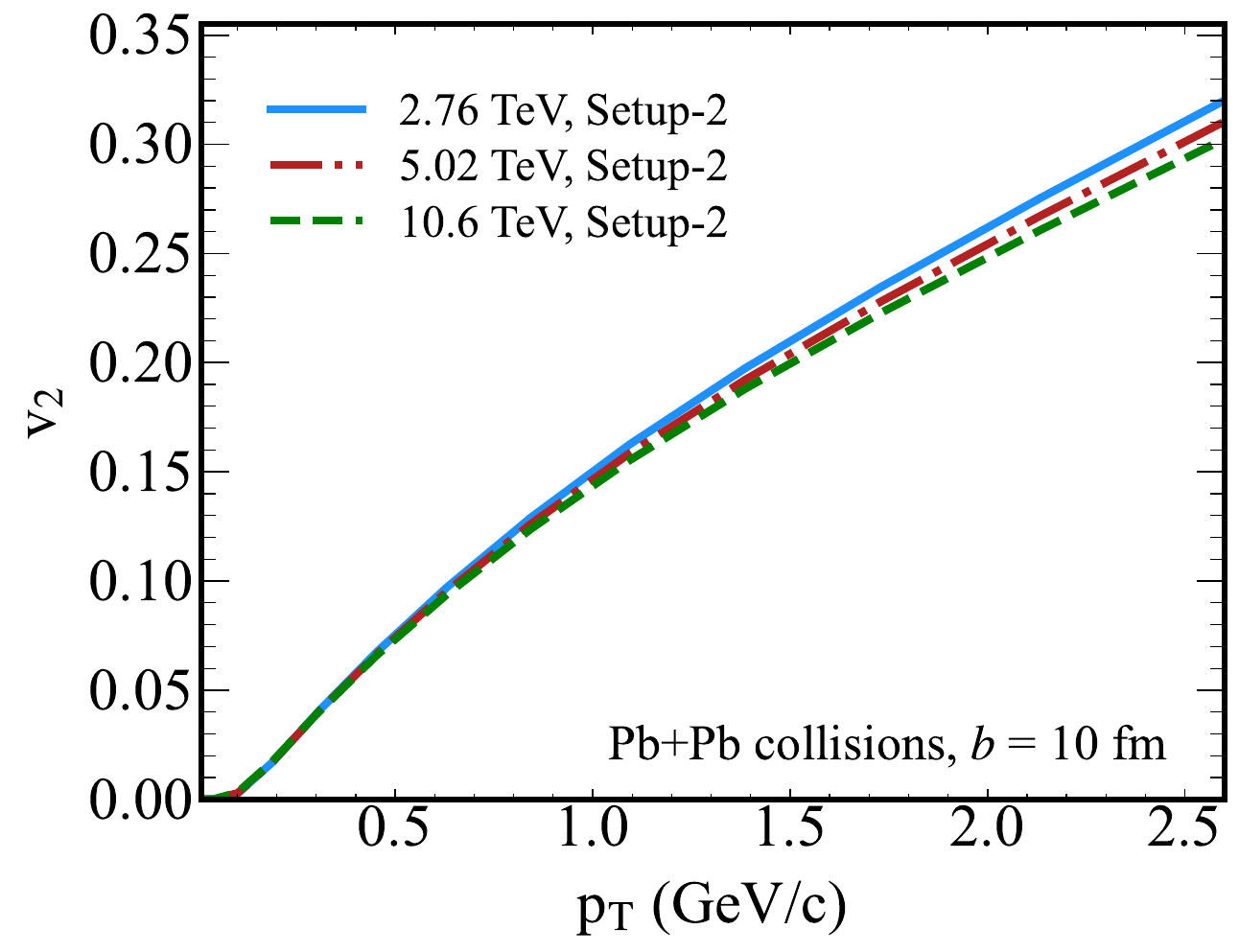}
\caption{(Color online) Upper panel: Pseudorapidity distribution $dN_{ch}/d\eta$ of charged hadrons in Pb+Pb collisions at centrality 0-5\% (b=2.6~fm) with $\sqrt{s_\text{NN}}=$ 2.76~TeV, 5.02~TeV and 10.6~TeV. The experimental data are obtained from the ALICE collaboration at ~2.76~TeV~\cite{ALICE:2015bpk}, ALICE collaboration~at 5.02~TeV~\cite{ALICE:2016fbt} and the HL-LHC prediction at~10.6~TeV~\cite{Citron:2018lsq}. Lower panel: Elliptic flow $v_2$ as a function of transverse momentum $p_\text{T}$ in Pb+Pb collisions (b=10~fm) with 
magnetic field Setup-2 in $\snn=$ 2.76~TeV, 5.02~TeV and 10.6~TeV.}
\label{Fig_snn_v2}
\end{figure}  

Previous studies have shown that the magnetic field decays more slowly in vacuum at RHIC energy due to the smaller relative speed between colliding nuclei~\cite{Pang:2016vdc}. The ratio of the maximum squeezing force density between RHIC and LHC collisions follows a proportionality of $(3m_{\pi}^{2}/78m_{\pi}^{2})^2\approx 0.0014$ at $\tau_{0}=0.2$ fm. Therefore, the paramagnetic squeezing effect is considered to be negligible at RHIC energy. However, for Pb+Pb non-collisions at the HL-LHC energy, where the maximum magnetic field can reach approximately 165$m_{\pi}^{2}$ at $\tau_{0}=0.2$ fm, the squeezing effect may become significant and deserves further investigation.  

In the upper panel of Figure~\ref{Fig_snn_v2}, we show the pseudorapidity distribution $dN_{\text{ch}}/d\eta$ of final charged hadrons at centrality 0-5\% ($b=2.6$~fm) in Pb+Pb collisionsin at $\snn=$ 2.76 TeV, 5.02 TeV and 10.6 TeV. Our model calculation results are in good agreement with the corresponding experimental data from ALICE collaboration and the 
CERN Yellow Report result of HL-LHC. This agreement validates the reliability of our current theoretical framework at the LHC energy regions.

In the lower panel of Figure~\ref{Fig_snn_v2}, we present the elliptic flow $v_2$ as a function of the transverse momentum $p_\text{T}$ in Pb+Pb collisions ($b=10$~fm) at different center-of-mass energies ($\snn=$~2.76~TeV, 5.02~TeV, and 10.6~TeV) under magnetic field Setup-2. 
It is observed that with increasing collision energy, a higher initial magnetic field is generated, leading to a larger squeezing force density and a more significant suppression of elliptic flow. We find that the hadrons elliptic flow is suppressed most (about 3-4\%) in 10.6~TeV Pb+Pb collisions at the HL-LHC energy.  

\section{SUMMARY}
\label{sec:4}

In this paper, we study the paramagnetic squeezing effect on anisotropic expansion of QGP in non-central Pb+Pb collisions at the LHC by simulating (3+1)-dimensional hydrodynamic with external magnetic fields. Three different external magnetic field setups and the most up-to-date magnetic susceptibility from lattice QCD calculation are used to study the squeezing force density on the $v_{2}$ of final hadrons in heavy ion collisions. 

We find that the squeezing effect is influenced by the lifetime of the magnetic field and its spatial distribution. A significant squeezing effect is observed for long-lived and widespread external magnetic fields at the initial stage of heavy ion collisions. However, we find that the commonly used magnetic field Setup-2 generates a weak squeezing force density on the QGP fireball, resulting in only a minor suppression of the elliptic flow $v_{2}$ of final hadrons.
For magnetic field Setup-2, the momentum eccentricity averaged on the whole bulk matter reduces by 10\% at $\tau_{0}=0.2$ fm, while the elliptic flow $v_{2}$ of final charged hadrons only reduces by 2\%. The reduction of $v_2$ due to the squeezing effect implies that the elliptic flow measured in the experiment is somewhat smaller than what would result from only the geometry expansion of the plasma. The suppression of the QGP fireball also suggest that electromagnetic probes such as thermal photons~\cite{Tuchin:2013ie} and open heavy flavor production~\cite{Das:2016cwd} might be more sensitive to the paramagnetic nuclear matter, since they travels the whole QGP medium. Additionally, we also find that the squeezing effect becomes larger at HL-LHC energy region, because the squeezing force density increased due to the magnetic field increase as the beam energy is increased.

We note that the current work can be extended in many aspects. First, due to the non-zero electric conductivity of the QGP, it is possible that the induced local magnetic field is large and has a long lifetime, both at RHIC and LHC energies as well as in lower energy collisions~\cite{Roy:2015kma,Huang:2022qdn}. The paramagnetic squeezing effect for those regions could be very long lifetime and result in different anisotropic flow. Second, the electric susceptibility $\xi$ is recently presented by lattice QCD calculation~\cite{Endrodi:2023wwf}, thus, it becomes very interesting to investigate how the electric field induced by the magnetic field affect the QGP evolution and the corresponding anisotropic flow in pA~\cite{Zhao:2021vmu,Zhao:2020wcd,Sun:2023adv} and AA collisions at the LHC. Finally, the transport coefficient of the QGP is affected by the electromagnetic field, thus, the relativistic magnetohydrodynamics with non-zero shear viscosity and bulk viscosity needed to develop for more accurate studies~\cite{Pu:2016ayh}. Theoretical studies with these aspects will be advanced in the near future.

\begin{acknowledgements}
We thank Anping Huang, Jinfeng Liao, Gergely Endr\H{o}di, Yafei Shi and Sheng-Qin Feng for helpful comments.  
This work was supported by the National Natural Science Foundation of China (NSFC) under Grant No.~12305138,
Guangdong Major Project of Basic and Applied Basic Research No.~2020B0301030008. 
\end{acknowledgements}

\bibliographystyle{unsrt}
\bibliography{paper}

\end{document}